\documentstyle[11pt]{article}

\makeatletter
\@addtoreset{equation}{section}
\makeatother

\def\text#1{\mbox{\rm #1}}
\input{tcilatex}
\begin{document}

\author{\makebox[13cm][l]{{\bf Delia Ionescu}\thanks{%
Department of Mathematics, Technical University of Civil Engineering ,
Bucharest, Romania ; E-mail: dionescu@hidro.utcb.ro}}}
\title{\parbox{15cm}{{\LARGE {\bf Can the notion of a homogeneous
gravitational field be transferred from classical mechanics to
the Relativistic Theory of Gravity?}}}}
\date{}
\maketitle

\begin{abstract}
\parbox{13cm}{The generalization of the concept of homogeneous gravitational field from
Classical Mechanics was considered in the framework of Einstein's General
Relativity by Bogorodskii. In this paper, I look for such a generalization
in the framework of the Relativistic Theory of Gravitation. There exist a
substantial difference between the solutions in these two theories.
Unfortunately, the solution obtained according to the Relativistic Theory of
Gravitation can't be accepted because it doesn't fulfill the Causality
Principle in this theory. So, it remains {\it open} in RTG the problem of
finding a generalization of the classical concept of homogeneous
gravitational field.}
\end{abstract}

\smallskip

\medskip

\bigskip \hsize 160mm\vsize 230mm

\section{Introduction}

\smallskip

In Newton's Classical Mechanics (CM), the homogeneous gravitational field is
the gravitational field which, in every point, has the same gradient of the
potential. Such a field is produced by an infinite material plane with the
constant surface density of mass. In Section 3, is presented this field in
CM.

It's natural to ask if the classical concept of the homogeneous
gravitational field can be conserved in the Relativistic Theory of
Gravitation (RTG). The source of inspiration in the analysis of this problem
is the monograph by Bogorodskii [3], Section 17.

In his monograph, Bogorodskii considers the problem of finding the
gravitational field produced by a system of masses uniformly distriduted on
a plane, according to Einstein's General Relativity Theory (GRT). In Section
4, I present what Bogorodskii means by homogeneous gravitational field in
GTR. But, as we'll see, his solution has an unccountable singularity wich
appears without any physical explanation.

The problem of such a homogeneous gravitational field in RTG, has briefly
considered by E. So\'{o}s and by me in the paper [5], Section 3. In Section
5 of the present paper, I analysed in all details this problem. The solution
of the complet system of RTG's Eqs. for the considered problem, differs from
Bogorodskii's solution. The obtained solution is regular in its entire
domaine of definition but it can't be acceptable like a real gravitational
field with physical sense because it doesn't fulfill the Causality Principle
(CP) in RTG. So, it remains {\it open} the problem of finding this field
according to RTG.

In section 6, I show that if the solution obtained by me is not rejected
using CP, the velocity of some free test particles in the produced field,
overpass the velocity of light in vacuum.

\section{RTG's equations and the Causality Principle in RTG}

\smallskip

RTG was constructed by Logunov and his co-workers (see [1], [2]) as a field
theory of the gravitational field within the framework of Special Relativity
Theory (SRT). The Minkowski space-time is a fundamental space that
incorporates all physical fields, including gravitation. The line element of
this space is:

\hspace{5cm} 
\begin{equation}
d\sigma ^{2}=\gamma _{mn}(x)dx^{m}dx^{n},
\end{equation}
where $x^{m},$ $m=$1, 2, 3, 4, is an admissible coordinate system in the
underlying Minkowski space-time; $\gamma _{mn}(x)$ are the components of the
Minkowskian metric in the assumed coordinate system.

The gravitational field is described by a second order symmetric tensor $%
\phi ^{mn}(x)$, owing to the action of which un effective Riemannian
space-time arises.

One of the basic assumption of RTG\ tells us that the behaviour of matter in
the Minkowskian's space -time with metric $\gamma _{mn}(x)$, under the
influence of the gravitational field $\phi ^{mn}(x),$ is identical to its
behaviour in the effective Riemannian space-time with metric $g_{mn}(x)$,
determined according to the rules:

\begin{equation}
\tilde{g}^{mn}=\sqrt{-g}\mbox{\rm  }g^{mn}=\sqrt{-\gamma }\mbox{\rm  }\gamma
^{mn}+\sqrt{-\gamma }\mbox{\rm  }\phi ^{mn}\mbox{\rm  },\mbox{\rm  }g=\det
(g_{mn})\mbox{\rm  },\mbox{\rm  }\gamma =\det (\gamma _{mn}).
\end{equation}

Such interaction of the gravitational field with matter was termed the
geometrisation principle of RTG.

The behaviour of the gravitational field is governed by the following
differential laws of RTG:

\begin{equation}
R_n^m-\frac 12\delta _n^mR+\frac{m_g^2}2\left( \delta _n^m+g^{mk}\gamma
_{kn}-\frac 12\delta _n^mg^{kl}\gamma _{kl}\right) =8\pi T_n^m,
\end{equation}

\begin{equation}
D_m\tilde{g}^{mn}=0\mbox{\rm  , \quad }m,n,k,l=1,2,3,4.
\end{equation}

Here $R_n^m$ is Ricci's tensor corresponding to $g_{mn}$ , $R=R_m^m$ is the
scalar curvature$,$ $\delta _n^m$ are Kronecker's symbols, $m_g$ is the
graviton mass and $T_n^m$ denotes the energy-momentum tensor of the sources
of the gravitational field. In (2.4) $D_m$ is the operator of covariant
differentiation with respect to the metric $\gamma _{mn}$. Eqs. (2.3), (2.4)
are covariant under arbitrary coordinate transformations with a nonzero
Jacobian. In RTG all field variables depend on the universal
spatial-temporal coordinates in the Minkowski space-time. The presence of
the mass terms in Eqs. (2.3) makes it possible to unambiguously determine
the geometry of space-time and the gravitational field energy-momentum
density in the absence of matter. Eqs. (2.4) tell us that a gravitational
field can have only the spin states 0 and 2. In the work [2], these Eqs.
which determine the polarization states of the field, are consequences of
the fact that the source of the gravitational field is the universal
conserved density of the energy-momentum tensor of the entire matter
including the gravitational field. The graviton mass substantially
influences on the Uniserse evolution and alters the nature of gravitational
collapse.

In this work, because of the extreme smallness of the graviton mass ($%
m_g\simeq 10^{-66}$ grams), we will analyse the problem of finding the
homogeneous\ gravitational field in RTG, considering Eqs. (2.3) without mass
terms.

Relativistic units are used in all Eqs. .

Eqs. (2.4) can be written in the following form (see [1], Appendix 1):

\begin{eqnarray}
D_{m}\tilde{g}^{mn}=\tilde{g}^{mn},_{m}+\gamma _{mp}^{n}\tilde{g}^{mp}=0, 
\end{eqnarray}
where $\gamma _{mp}^{n}$ are the components of the metric connection
generated by $\gamma _{mn}$ and the comma is the derivation relative to the
involved coordinate. The causality principle (CP) in RTG is presented and
analysed by Logunov in [2], Section 6.

According to CP any motion of a pointlike test body must have place within
the causality light cone of Minkowski's space-time. According to Logunov's
analysis CP will be satisfied if and only if for any isotropic Minkowskian
vector $u^{m}$ , i.e. for any vector $u^{m}$ satisfying the condition:

\begin{equation}
\gamma _{mn}u^{m}u^{n}=0,
\end{equation}
the metric of the effective Riemannian space-time satisfies the restriction:

\begin{equation}
g_{mn}u^{m}u^{n}\leq 0
\end{equation}

According to CP of RTG\ only those solutions of the system (2.3), (2.4) can
have physical meaning which satisfies the above restriction.

It's important to stress the fact that CP\ in the above form can be
formulated only in RTG, because only in this theory, the space-time is
Minkowskian and the gravitational field is described by a second order
symmetric tensor field $\phi _{mn}(x)$, $x^{m}$ being the admissible
coordinates in the underlying Minkowskian space-time, $x^{1},x^{2},x^{3}$
being the space-like variables and $x^{4}$ being the time-like variable.

\section{Homogeneous\ gravitational field in CM}

\smallskip

In CM a gravitational field is named homogeneous if its intensity is a
constant magnitude or a picewise constant magnitude. Such a field is
generated by a system of masses uniformly distributed on a plane. The
connection between the surface density $\sigma $ of the mass and the
acceleration ${\cal G}$ due to the produced gravitational field is given by
the relation:

\begin{equation}
{\cal G}=2\pi \sigma k>0,
\end{equation}
$k$ being Newton's gravitational constant.

Choosing the Cartesian axes $x$ and $y$ in the mentioned plane and the axis $%
z$ perpendicular to this plane, the motion of a free test particle in this
gravitational field is governed by Eqs.:

\begin{equation}
\frac{d^{2}x}{dt^{2}}=0\mbox{\rm  , }\frac{d^{2}y}{dt^{2}}=0\mbox{\rm  }
\end{equation}

\begin{equation}
\frac{d^{2}z}{d^{2}t}+{\cal G}=0\mbox{\rm  , for }z>0%
\mbox{\rm \quad and
\quad }\frac{d^{2}z}{dt^{2}}-{\cal G}=0\mbox{\rm  , for }z<0,
\end{equation}
where $t$ is the Newtonian time.

It can be also notice that in a non-inertial frame which moves with constant
proper acceleration ${\cal G,}$ the laws of motion of a free test particle
have the following form:

\begin{equation}
\frac{d^{2}x}{dt^{2}}=0\mbox{\rm  , }\frac{d^{2}y}{dt^{2}}=0\mbox{\rm  , }%
\frac{d^{2}z}{dt^{2}}+{\cal G}=0\mbox{\rm  , for any }z.
\end{equation}
Here $x,$ $y,$ $z$ represent the coordinates of the particle in the
non-inertial frame.

In spite of the formal similarity between the system of Eqs. (3.2), (3.3)
and Eqs. (3.4), these two sets of laws are not equivalent. For this, it's
sufficient to compare Eqs. (3.3) and (3.4)$_{3}.$ While the laws (3.4) can
take, by a transformation of frame, the following form:

\begin{equation}
\frac{d^{2}X}{dt^{2}}=0\mbox{\rm  , }\frac{d^{2}Y}{dt^{2}}=0\mbox{\rm  , }%
\frac{d^{2}Z}{dt^{2}}=0\mbox{\rm  , for any }Z,
\end{equation}
such an operation it's not possible for the laws (3.2), (3.3). The
non-equivalence between these two sets of laws reflects the fact that in the
first case we deal with a real gravitational field unlike the second case in
which an inertial force is present.

That example shows that even in CM, there is a difference between a real
gravitational force produced by a distribution of mass and an inertial force
acting in a non-inertial frame.

\smallskip

\section{Homogeneous\ gravitational field in GTR}

\smallskip

\smallskip In his monograph [3], Section 17, Bogorodskii has searched for an
answer to the following questions: There exists such a homogeneous
gravitational field in GTR? Which is the Riemannian metric that represents
in GTR the gravitational field produced by an infinite material plane with
constant surface density of mass? His answers at these questions are
presented below.

Taking into account the classical results, Bogorodskii looks for the
solutions of Einstein's Eqs. in the form:

\begin{equation}
ds^{2}=-A\mbox{\rm  }dx^{2}-A\mbox{\rm  }dy^{2}-C\mbox{\rm  }dz^{2}+D%
\mbox{\rm  }dt^{2}
\end{equation}
\ where $A$, $C$, $D$ are positive functions depending only on $z$.

The energy -momentum tensor of the sources which are uniformely distributed
on the plane $z$=0 is:

\begin{equation}
T^{mn}\equiv 0,\mbox{\rm  for any }z\neq 0.
\end{equation}

For the metric (4.1) and for the expression of the energy-momentum tensor
(4.2), Bogorodskii concludes that Einstein's field Eqs. are fulfilled if the
unknown functions $A,$ $C,$ $D,$ satisfy the following Eqs.:

\[
2\left( \frac{A^{^{\prime }}}{A}\right) ^{^{\prime }}-\frac{A^{^{\prime
}}C^{^{\prime }}}{AC}+\frac{A^{^{\prime }}}{A}\left( \frac{2A^{^{\prime }}}{A%
}+\frac{D^{^{\prime }}}{D}\right) =0, 
\]

\begin{equation}
2\left( \frac{D^{^{\prime }}}{D}\right) ^{^{\prime }}-\frac{C^{^{\prime
}}D^{^{\prime }}}{CD}+\frac{D^{^{\prime }}}{D}\left( \frac{2A^{^{\prime }}}{A%
}+\frac{D^{^{\prime }}}{D}\right) =0,
\end{equation}

\[
\frac{A^{^{\prime }}}{A}\left( \frac{A^{^{\prime }}}{A}+\frac{2D^{^{\prime }}%
}{D}\right) =0. 
\]
Here the prime mark denotes the derivation relative to the coordinate $z.$

According to the last Eq., there are two possibilities:

\begin{equation}
A^{^{\prime }}=0\mbox{\rm  \quad or \quad }\left( \frac{A^{^{\prime }}}{A}+%
\frac{2D^{^{\prime }}}{D}\right) =0.
\end{equation}

In the first case it can be taken $A=1,$ because the functions $A,$ $C,$ $D$
are determined up to a constant. In this case, Eq. (4.3)$_{1}$ is evidentely
fulfilled and (4.3)$_{2}$ becomes:

\begin{equation}
\left( \frac{D^{^{\prime }}}{D}\right) ^{^{\prime }}-\frac{1}{2}\frac{%
C^{^{\prime }}D^{^{\prime }}}{CD}+\frac{1}{2}\left( \frac{D^{^{\prime }}}{D}%
\right) ^{2}=0.
\end{equation}

This yields $C=aD^{-1}D^{^{\prime }}$ $^{2}$, $a$ being a real constant of
integration.

So, for this case the solution of Einstein's Eqs. has the form:

\begin{equation}
A=1\mbox{\rm  , \quad }C=aD^{-1}D^{^{\prime }}\mbox{\rm  }^{2},
\end{equation}
$D$ being an arbitrary function on $z$ .

For the second case it can be taken $A=D^{-2}$ and Eqs. (4.3)$_{1}$, (4.3)$%
_{2}$ become:

\begin{equation}
\left( \frac{D^{^{\prime }}}{D}\right) ^{^{\prime }}-\frac{1}{2}\frac{%
C^{^{\prime }}D^{^{\prime }}}{CD}-\frac{3}{2}\left( \frac{D^{^{\prime }}}{D}%
\right) ^{2}=0.
\end{equation}

This yields $C=bD^{-5}D^{\prime }$ $^{2}$, $b$ being a constant of
integration.

Thus for the second case, the solution of Einstein's Eqs. has the form:

\begin{equation}
A=D^{-2}\mbox{\rm  , }C=bD^{-5}D^{\prime }\mbox{\rm  }^{2},
\end{equation}
$D$ being an arbitrary function on $z$ .

With the view of finding $D$($z$), Bogorodskii observes that the motion of a
free test particle in the produced gravitational field is determined by Eqs.
of geodesics:

\begin{equation}
\frac{d^{2}x}{dt^{2}}+\left( \frac{A^{\prime }}{A}-\frac{D^{\prime }}{D}%
\right) \frac{dx}{dt}\frac{dz}{dt}=0\mbox{\rm  , }\frac{d^{2}y}{dt^{2}}%
\mbox{\rm  }+\left( \frac{A^{\prime }}{A}-\frac{D^{\prime }}{D}\right) \frac{%
dy}{dt}\frac{dz}{dt}=0\mbox{\rm  ,}
\end{equation}

\begin{equation}
\frac{d^{2}z}{dt^{2}}-\frac{A^{\prime }}{2C}\left[ \left( \frac{dx}{dt}%
\right) ^{2}+\left( \frac{dy}{dt}\right) ^{2}\right] +\left( \frac{C^{\prime
}}{2C}-\frac{D^{\prime }}{D}\right) \left( \frac{dz}{dt}\right) ^{2}+\frac{%
D^{\prime }}{2C}=0.
\end{equation}

From the above system it can be seen that the vertical motion, with the
velocity equals to zero at the initial moment, is described by the system of
Eqs.:

\begin{equation}
\frac{d^{2}x}{dt^{2}}=0\mbox{\rm  , }\frac{d^{2}y}{dt^{2}}=0\mbox{\rm  ,}
\end{equation}

\begin{equation}
\frac{d^{2}z}{dt^{2}}+\left( \frac{C^{\prime }}{2C}-\frac{D^{\prime }}{D}%
\right) \left( \frac{dz}{dt}\right) ^{2}+\frac{D^{\prime }}{2C}=0. \ 
\end{equation}

In the case of slow motion, the term which contains the velocity must be
omitted, so, Eq.(4.12) becomes : 
\begin{equation}
\frac{d^{2}z}{dt^{2}}+\frac{D^{\prime }}{2C}=0.
\end{equation}

Comparing the system of Eqs. (4.11), (4.13) with the classical one,
Bogorodskii requires:

\begin{equation}
\frac{D^{\prime }}{2C}={\cal G}.
\end{equation}

I'll return at this condition.

Finally, taking the constants $a,$ $b$ equal to $\frac{1}{4{\cal G}^{2}}$
and using the relation (4.14), from (4.6), (4.8) Bogorodskii obtaines the
solutions:

\begin{equation}
A=1\mbox{\rm  , }C=e^{2{\cal G}z}\mbox{\rm  , }D=e^{2{\cal G}z}\mbox{\rm  ,}
\end{equation}
and

\begin{equation}
A=(1-8{\cal G}z)^{1/2}\mbox{\rm  , }C=(1-8{\cal G}z)^{-5/4}\mbox{\rm  },%
\mbox{\rm  }D=(1-8{\cal G}z)^{-1/4}.
\end{equation}

For the first solution (4.15), the Riemann-Christoffel curvature tensor is
identically zero. Thus, the autor concludes that (4.15) does not represent a
real gravitational field. It's easy to see that by the transformations:

\begin{equation}
X=x\mbox{\rm  , }Y=y\mbox{\rm  , }Z=\frac{1}{{\cal G}}[e^{{\cal G}z}ch({\cal %
G}t)-1]\mbox{\rm  , }T=\frac{1}{{\cal G}}e^{{\cal G}z}sh({\cal G}t)%
\mbox{\rm
,}
\end{equation}
the fundamental invariant (4.1) becomes the Minkowskian one:

\begin{equation}
d\sigma ^{2}=-dX^{2}-dY^{2}-dZ^{2}+dT^{2}.
\end{equation}

The properties and the singularities of the non-inertial frame characterized
by the relations (4.17) are studied in detail in the monograph of Jukov [4],
Section15. I don't present here these properties.

Consequently, the first solution (4.15) corresponds to a non-inertial frame
whose origine moves with the constant proper acceleration ${\cal G}$ along
the pozitive axis Z of an inertial frame.

The Riemann-Christoffel curvature tensor corresponding to the second
solution (4.16) is not zero. According to Bogorodskii, this solution
represents the real homogeneous gravitational field in GRT, produced by the
considered distribution of mass.

First of all, it can be observed that the solution (4.16) has a strange
singularity in $z=\frac{1}{8{\cal G}}$ , which is difficult to be explained.

Now, it's the moment to return to the condition (4.14) required by
Bogorodskii. It has been pointed out in Section 3, that in CM the motion of
a free test particle in the real gravitational field is governed by the
system of Eqs. (3.2), (3.3) not by the system (3.4). Thus, Bogorodskii's
relation (4.14) must be replaced with:

\begin{equation}
\frac{D^{\prime }}{2C}=\left\{ 
\begin{array}{c}
{\cal G}\mbox{\rm  \quad , }z>0 \\ 
-{\cal G}\mbox{\rm  \quad , }z<0
\end{array}
\right.
\end{equation}

Hence, Bogorodskii's solution (3.12) must be replaced with:

\begin{equation}
A=(1\mp 8{\cal G}z)^{1/2}\mbox{\rm  , }C=(1\mp 8{\cal G}z)^{-5/4}%
\mbox{\rm 
, }D=(1\mp 8{\cal G}z)^{-1/4}\mbox{\rm  ,}
\end{equation}
the sign + corresponds to $z<0$, the sign - corresponds to $z>0$.

The solution (4.20) can be accepted only for $-\frac{1}{8{\cal G}}<z<\frac{1%
}{8{\cal G}}$ and the singularities in $z=\pm \frac{1}{8{\cal G}}$ still
remain without any physical explanation.

\smallskip

\section{Homogeneous\ gravitational field in RTG}

\smallskip

The problem of finding the homogeneous gravitational field according to RTG
was shortly considered by E. So\'{o}s and by me in the paper [5], Section 3.
I present here all details concerning this problem.

To study any problem in RTG's framework, one must solve Eqs. (2.3), (2.4) in
terms of the coordinates of the underlying Minkowski space-time. Only those
solutions that satisfiy CP could represent physical acceptable gravitational
fields.

I keep the same point of departure as Bogorodskii. So, I look for the
solutions of this problem in RTG in the form (4.1). Two approaches can be
employed in obtaining these solutions. The both will be presented.

At first, I use the already found solutions, (4.15), (4.20), which fulfill
Eqs. (2.3). I verify if these solutions satisfy Eqs. (2.4).

For the solution (4.15), the components of the underlying Minkowskian metric
and the components of the Riemannian effective metric coincide because this
solution appears at the transition from an inertial reference system to an
accelerated reference system in Minkowski's space-time.Thus, Eqs. (2.4) are
obviously satisfied, $D_{m}$ being the operator of covariant differentiation
with respect to the Minkowskian metric.

For the solution (4.20) , first of all, it must be establish the reference
system of the underlying Minkowski space-time. This reference system can be
obtained assuming the vanishment of the gravitational field. In this way,
for $\sigma $=0 and implicitly ${\cal G}$=0, the metric (4.1) in the above
reference system becomes:

\begin{equation}
d\sigma ^{2}=-dx^{2}-dy^{2}-dz^{2}+dt^{2}.
\end{equation}

Consequently, for the chosen system of coordinates, the components of the
metric connection $\gamma _{mp}^{n}$ are zero and Eqs. (2.5) have the
simple form:

\begin{equation}
\tilde{g}^{mn},_{m}=0\mbox{\rm .}
\end{equation}

Taking into account (2.2), (4.1), (4.20), Eqs. (5.2) are not fulfilled. So,
(4.20) is not an admissible solution in RTG. For finding an admissible
solution in RTG, I use the same procedure as Logunov and Mestvirishvili in
[1], Chapter 13. Thus, I am looking for a system of coordinates \{$\eta ^{i}$%
\}$=\{X,$ $Y,$ $Z,$ $T\}$ in which Eqs. (2.3) are fulfilled, Eqs. (2.4)
establishing a one to one relationship between the sets of coordinates \{$%
\eta ^{i}$\} and \{$\xi ^{i}$\}=\{$x,$ $y,$ $z,$ $t$\} in the Minkowski
space-time. This change is made in such a way that, when the gravitational
field is swiched off, we arrive in the Minkowski space-time with Galilean
metric:

\begin{equation}
d\sigma ^{2}=-dX^{2}-dY^{2}-dZ^{2}+dT^{2}.
\end{equation}

Thus, $\gamma _{mp}^{n}$ in this system of coordinates are identically null.
Because the components of the metric (4.1) depend only on $z$, I shift from
the variables \{$\xi ^{i}$\} to the variables \{$\eta ^{i}$\} assuming that:

\begin{equation}
X=x,\mbox{\rm  }Y=y,\mbox{\rm  }Z=Z(z),\mbox{\rm  }T=t.
\end{equation}

We write Eqs. (2.4) in the chosen system of coordinates, in a somewhat
different form (see the relations (13.17), (13.22), from [1]):

\begin{equation}
\frac{\partial }{\partial \xi ^{m}}\left( \sqrt{-g(\xi )}g^{mn}(\xi )\frac{%
\partial \eta ^{p}}{\partial \xi ^{n}}\right) =0.
\end{equation}

For the transformation (5.4), taking into account (4.1), (4.20), the system
(5.5) becomes:

\begin{equation}
\frac{d}{dz}\left( \left( 1\mp 8{\cal G}z\right) \frac{dZ}{dz}\right) =0.
\end{equation}

Integrating this Eq. and choosing the constant of integration, in such a way
that for ${\cal G}$ converges to zero, $Z$ converges to $z,$we get:

\begin{equation}
Z=\mp \frac{1}{8{\cal G}}\ln \left( 1\mp 8{\cal G}z\right) .
\end{equation}

Now, it's sufficient to employ the tensor transformation law and we find the
components of the metric (4.1), in this second case, in the the system of
coordinates (5.4), (5.7):

\[
A=e^{-4{\cal G}Z}\mbox{\rm  , }C=e^{-6{\cal G}Z}\mbox{\rm  , }D=e^{2{\cal G}%
Z}\mbox{\rm , for any }Z>0 
\]

\begin{equation}
A=e^{4{\cal G}Z}\mbox{\rm  , }C=e^{6{\cal G}Z}\mbox{\rm  , }D=e^{-2{\cal G}Z}%
\mbox{\rm , for any }Z<0
\end{equation}
The solution (5.8) satisfies the complet system of Eqs. (2.3), (2.4). This
solution is regular for any $Z\neq 0.$ Anyway, it is not derivable for $Z$=0
and crossing this plane the derivatives of the functions $A,$ $C,$ $D$ have
finite jumps. Of course, the appearence of this singularity, concentrated in
the plane $Z$=0, reflects the fact that this real gravitational field has
like source a system of masses distributed on the respective plane. We
observe also that the condition (4.19) can be only approximately fulfilled,
because, for example, from (5.8)$_{1}$:

\begin{equation}
\frac{D^{^{\prime }}}{2C}={\cal G}\mbox{\rm  }e^{8{\cal G}Z}%
\mbox{\rm , for
any }Z>0.
\end{equation}

As the same time, we notice that this approximation is justified somehow. If
we use the usual units, we get:

\begin{equation}
\frac{D^{^{\prime }}}{2C}=\frac{{\cal G}}{c^{2}}\mbox{\rm  }e^{\frac{8{\cal G%
}Z}{c^{2}}}\mbox{\rm, for any }Z>0,
\end{equation}
$c$ being the speed of light in vacuum, relative to an inertial frame.
Consequently, for any $Z>0$ with:

\begin{equation}
Z\ll \frac{c^{2}}{{\cal G}},
\end{equation}
the ratio $\frac{D^{^{\prime }}}{2C}$ can be considered approximately
constant.

The analysis in RTG can't be stoped here, since the solutions must be also
submitted to CP.

For the first solution (4.15), CP is fulfilled evidently, because in this
case, as it was already mentioned, the Minkowskian metric and the Riemannian
metric are the same. Thus, the solution (4.15) is also an admissible
solution in RTG. Its physical significance was already clarified.

For the second solution (5.8), taking into account the form (5.3) of the
underlying Minkowski space-time, $u$=(1, 0, 0,1) is an isotropic Minkowskian
vector. Consequently, the condition(2.7) is fulfilled if:

\begin{equation}
e^{2{\cal G}Z}\leq e^{-4{\cal G}Z}\mbox{\rm  , for }Z>0%
\mbox{\rm \quad and
\quad }e^{-2{\cal G}Z}\leq e^{4{\cal G}Z}\mbox{\rm  , for }Z<0.
\end{equation}

These conditions are not fulfilled for any $Z>0$ or $Z<0$, if ${\cal G}$ is
not zero.

We conclude that in accordance with RTG, can't exist a generalization of the
homogenouse gravitational field in Bogorodskii's sense.

It can also be employed the following approache for finding the solution of
the considered problem in RTG. From Eqs. (2.3) with $T_{n}^{m}\equiv 0,$ for 
$z\neq 0,$ we get the solutions (4.6), (4.8). I stress that in (4.6), (4.8), 
$D(z)$ is an arbitrary function. This fact shows clearly that Einstein's
field Eqs., are not enough for finding in an unique manner the gravitational
field produced by the considered distridution of mass. I determine the
unknown function $D(z)$, using Eqs. (2.4).

Let us consider $x$, $y$, $z$, $t$ the Galilean coordinates of an inertial
frame. So, Eqs. (2.4) take the simple forme (5.2). Taking into account
(2.2), (4.1), (4.6), (4.8), from (5.2) we get:

\begin{equation}
D(z)=pe^{qz}\mbox{\rm  ,}
\end{equation}
where $p$ and $q$ are real constants.

Thus, introducing (5.13) into (4.6), we obtain the first solution according
to RTG:

\begin{equation}
A=1\mbox{\rm  , }C=ape^{qz}\mbox{\rm  , }D=pe^{qz}
\end{equation}
and from (4.8) the second solution:

\begin{equation}
A=p^{-2}e^{-2qz}\mbox{\rm  , }C=bp^{-3}e^{-3qz}\mbox{\rm  , }D=pe^{qz}%
\mbox{\rm  .}
\end{equation}

For the first solution (5.14), the Riemann-Christoffel curvature tensor is
identically zero and for the second solution (5.15) it's different from zero.

The constants $a$, $b$, $p$, $q$ must be determined from the Correspondence
Principle: after switching off the gravitational field, the curvature of
space disappears and we find ourselves in the Minkowski space-time in the
chosen reference system. So, Eqs. of motion become classical Eqs. of motion
in the chosen reference system. From the geometrization principle of RTG,
Eqs. of motion in the considered gravitational field are given by Eqs.
(4.9), (4.10). In the case of vertical and slow motion, this system of Eqs.
become the system (4.11), (4.13). Thus, from the Correspondence Principle,
we must have the relation (4.14) for the solution (5.14) and the relation
(4.19) for the solution (5.15).

For the same principle, the metric must tend to the Galilean metric for $%
{\cal G}$ converges to zero. Thus, in the first case, we get the solution
(4.15) and in the second case the solution (5.8). For this last solution the
relation (4.19) is only approximately fulfilled, as we have already discused.

So, by the two approaches, we have obtained the same result: one solution,
(4.15), that represents an inertial force and the other, (5.8), which can't
be accepted as the real homogeneous gravitational field produced by the
considered distribution of mass, because it doesn't fulfill CP (see 5.12).

\smallskip

\section{The importance of CP in RTG}

\smallskip

The example which will be presented in this Section shows clearly the
importance of CP in deciding the rejection of the solution (5.8). I'll show
that there exist free test particles, in the obtained space-time (5.8),
which move quickly than the light in vacuum.

Let us consider a free test particle, situated at the initial moment $t=0$,
at the distance $h>0$ from the plane $z$=0, where the masses are
concentrated. For the sake of simplicity, we consider the problem in the
plane $xOz.$ At the initial moment $t$=0, when the particle has the assumed
position:

\begin{equation}
x(0)=0
\end{equation}

\begin{equation}
z(0)=h>0
\end{equation}
it is thrown up.

From the geometrization principle, if we want to study the behavior of this
particle under the influences of the considered homogeneous gravitational
field, we can study its behaviour in the following effective Riemannian
space-time (see (5.8)$_{1}$):

\begin{equation}
ds^{2}=-e^{-4{\cal G}z}dx^{2}-e^{-6{\cal G}z}dz^{2}+e^{2{\cal G}z}dt^{2}%
\mbox{\rm  }
\end{equation}

Taking into account Eqs. (4.9)$_{1}$, (4.10) of geodesics and the expression
(6.3) of the Riemannian metric, the trajectories of the particle are
described by: 
\begin{equation}
\frac{d^{2}x}{dt^{2}}-6{\cal G}\frac{dx}{dt}\frac{dz}{dt}=0\mbox{\rm  ,}
\end{equation}

\begin{equation}
\frac{d^{2}z}{dt^{2}}+2{\cal G}e^{2{\cal G}z}\left( \frac{dx}{dt}\right)
^{2}-5{\cal G}\left( \frac{dz}{dt}\right) ^{2}+{\cal G}e^{8{\cal G}z}=0.
\end{equation}

We also consider that at the initial moment $t$=0:

\begin{equation}
\dot{x}(0)=a>0,
\end{equation}

\begin{equation}
\dot{z}(0)=b>0,
\end{equation}
$a,$ $b$ beaing real constants.

Solving Eq. (6.4) and taking into account (6.2), (6.6), we get:

\begin{equation}
\dot{x}(t)=ae^{6{\cal G}(z-h)}.
\end{equation}

Also solving Eq. (6.5), on the basis of (6.2), (6.6), (6.7), we get:

\begin{equation}
\dot{z}(t)=e^{4{\cal G}z}\sqrt{1+Le^{2{\cal G}z}-a^{2}e^{-12{\cal G}h}e^{6%
{\cal G}z}.}
\end{equation}
Here $L$ is a real constant equals to:

\begin{equation}
L=b^{2}e^{-10{\cal G}h}+a^{2}e^{-8{\cal G}h}-e^{-2{\cal G}h}.
\end{equation}

Now, we single out in (6.3) the time-like part and the space-time part:

\begin{equation}
ds^{2}=d\sigma ^{2}-dl^{2},
\end{equation}
where:

\begin{equation}
d\sigma ^{2}=e^{2{\cal G}z}dt^{2},
\end{equation}

\begin{equation}
dl^{2}=e^{-4{\cal G}z}dx^{2}+e^{-6{\cal G}z}dz^{2}.
\end{equation}

So, if a pointlike event has coordinates ($x,$ $0,$ $z,$ $t$) and another
pointlike event has coordinates ($x+dx,$ $0,$ $z+dz,$ $t+dt$), then an
observer in ($x,$ $0,$ $z,$ $t$) with four-velocity ($\frac{dx}{ds},$ $0,$ $%
\frac{dz}{ds},$ $\frac{dt}{ds}$), measures between the two events a proper
spatial distance $dl$ and an interval of proper time $d\sigma .$

The velocity $v$($v^{1}=\frac{dx}{dt},$ $0,$ $v^{3}=\frac{dz}{dt}$) of the
particle, in the considered effective Riemannian space-time, has the
following absolute value:

\begin{equation}
v^{2}=\frac{dl^{2}}{dt^{2}}=e^{-4{\cal G}z}\dot{x}^{2}+e^{-6{\cal G}z}\dot{z}%
^{2}
\end{equation}

It is natural to demand that the initial velocity of our particle be smaller
than the velocity of light in vacuum. So, from (6.2), (6.6), (6.7), (6.14),
we get:

\begin{equation}
e^{-4{\cal G}h}a^{2}+e^{-6{\cal G}h}b^{2}<1.
\end{equation}

Taking into account (6.10), (6.15), we obtain the following restriction for
the constant $L:$

\begin{equation}
L<e^{-4{\cal G}h}-e^{-2{\cal G}h}
\end{equation}

We have considered $h>0$, so, from (6.16):

\begin{equation}
L<0.
\end{equation}

Now, I show that for the case chosen by me, there are some coordinates $z$,
such that the velocity of the particle at these points overpass the velocity
of light in vacuum.So, I find some coordinate $z$ such that:

\begin{equation}
v^{2}(z)>1.
\end{equation}

Introducing (6.8), (6.9) into (6.14), the inequality (6.18) is equivalent to:

\begin{equation}
e^{2{\cal G}z}+Le^{4{\cal G}z}>1.
\end{equation}

Taking into account (6.17), by some calculus, we obtain that for:

\begin{equation}
-\frac{1}{4}<L<0,
\end{equation}
so, for some conditions imposed on the initial values of the velocity, for
any $z$ satisfying:

\begin{equation}
\frac{1}{2{\cal G}}\ln \left( \frac{-1+\sqrt{1+4L}}{2L}\right) <z<\frac{1}{2%
{\cal G}}\ln \left( \frac{-1-\sqrt{1+4L}}{2L}\right) ,
\end{equation}
the inequality (6.19) is fulfilled.

For example, the restrictions (6.20) are fulfilled for the following $a$ and 
$b$:

\begin{eqnarray*}
a &=&\rho e^{2{\cal G}h}\cos \theta , \\
b &=&\rho e^{3{\cal G}h}\sin \theta ,\mbox{\rm  with }\theta \in \left(
0,\pi /2\right) ,\mbox{\rm  }0<\rho <1,\mbox{\rm  }\rho ^{2}>1-\frac{(e^{^{2%
{\cal G}h}}-2)^{2}}{4}.
\end{eqnarray*}
Hence, if we consider in the effective Riemannian space-time (5.8)$_{1}$ a
free test particle, which has at the initial time the position (6.1), (6.2)
and the velocity (6.6), (6.7), such that, taking into account (6.10), the
real constants $a,$ $b$ satisfy the restrictions (6.20), this particle moves
quikly than the light in vacuum.

\section{Conclusions}

\smallskip

As we have seen, in CM, GRT, RTG, if a frame is moving at a constant proper
acceleration, relative to an inertial frame, then the inertial field due to
the force of inertia, is a constant field. The expression of the Minkowski
line element (4.15) is the same in GRT and RTG. In CM, it was considered
that this constant field is indistingushable from a homogeneous
gravitational field produced by an infinite material plane. But from (3.2),
(3.3) and (3.4) we have seen that there exists a difference, even in CM,
between these two fields. This is due to the fact that the gravitational
force is a force of attraction. In GRT and RTG the difference between the
constant field produced by an inertial force and the homogeneous
gravitational field due to the presence of mass, is substantial.
Bogorodskii's solution in GRT, for the homogeneous gravitational field, is
given by (4.16). As we have already discused, his solution has an
unccountable singularity. The solution obtained in RTG has the form (5.8).
Unfortunately, this solution can't be kept because it doesn't satisfy CP in
RTG. The decision of the rejection of this solution, using CP, it's right.
Indeed, in the obtained space-time (5.8), the velocity of a free test
particle can overpass the velocity of light in vacuum. Ending to this
analysis, I can conclude that the problem of finding in RTG the
gravitational field produced by a uniform distribution of mass concentrated
on an infinite plane is a very interesting problem, but which remains {\it %
open}.

\smallskip \smallskip {\bf Acknowledgements. }I would like to thank my
teacher E. So\'{o}s, for many helpful and interesting discussions which
stimulated me to undertake this work. I am also grateful to Prof. V. Soloviev
for translating this paper into Russian.

\end{document}